\documentclass{elsart}

\def\la{\raise0.3ex\hbox{$<$}\kern-0.75em{\lower0.65ex\hbox{$\sim$}}}
\def\ga{\raise0.3ex\hbox{$>$}\kern-0.75em{\lower0.65ex\hbox{$\sim$}}}
\def\kms{\rm ~km~s^{-1}}
\def\Msun{~M_\odot}

\def\Ha{${\rm H}\alpha$}

\def\EE#1{\times 10^{#1}}

\def\wll{\lambda\lambda}
\def\coa{${}^{56}\rm Co$}
\def\cob{${}^{57}\rm Co$}
\def\ti{${}^{44}\rm Ti$}
\def\nib{${}^{57}\rm Ni$}
\def\nia {${}^{56}$Ni}
\def\isotope#1#2{\hbox{${}^{#1}\rm#2$}}
\def\apj{ApJ}
\def\apjl{ApJ Lett}
\def\aj{AJ}
\def\aap{A\&A}

\usepackage{natbib}

\usepackage{graphicx}

\usepackage{amssymb}

\begin{document}

\begin{frontmatter}



\title{Radioactivities and nucleosynthesis in SN 1987A}


\author{Claes Fransson and Cecilia Kozma}
\address{Stockholm Observatory, Department of Astronomy, SCFAB, SE-106 91 Stockholm}

\begin{abstract}
The nucleosynthesis and production of radioactive elements in SN 1987A
are reviewed. Different methods for estimating the masses of \nia,
\nib, and \ti~ are discussed, and we conclude that broad band
photometry in combination with time-dependent models for the light
curve gives the most reliable estimates.
\end{abstract}


\end{frontmatter}

\section{Introduction}
\label{sec_introd}
Two of the outstanding issues for SNe are the nucleosynthesis and the
explosion mechanism. Although SNe have been known from the 1950's to
be the most important sources of heavy elements, there is little
quantitative evidence for this. SN 1987A has in this respect been a
unique source of information thanks to the possibility to obtain
spectral information at very late epochs when the ejecta is
transparent and the central regions observable. The nature of the
explosion mechanism is still to a large extent unknown, and
constraints from observations are badly needed. The main such
constraints are provided by the hydrodynamic structure, e.g., the
extent of mixing, and the masses of the iron peak elements, in
particular the radioactive isotopes \citep[e.g.,][]{Timmes et al. (1996),
Kumagai et al. (1993), 2000ApJ...531L.123K}.

\section{Abundances}
To model the spectra and light curve a complicated chain of
thermalization processes from the gamma-rays and positrons in the keV
-- MeV range, to the observed optical and IR photons has to be
understood. The details of this are discussed in a series of papers,
\citep{kf92,kf98a,kf98b}. See also \citet{Fransson (1994)} and
\citet{deKool98}. Here we only summarize the main points.

After a couple of days the main energy input to the SN ejecta comes
from radioactive decay, first \isotope{56}{Ni} followed by
\isotope{56}{Co}.  Later than $\sim 1100$ days \isotope{57}{Co} takes
over, while at very late epochs, $\ga 2000$ days \ti~ decay becomes
dominant. Table 1 summarizes the different decay routes, the emitted
particle (gamma-rays or positrons), the exponential decay time scale
and the approximate epoch when the various decays dominated the energy
input to SN 1987A.

\begin{table}[t]
\caption{Main radioactive decays in SN 1987A}
\begin{tabular}{rlllrc}
\hline
Decay &&&& Time scale & Epoch when dominating\\ 
\hline
$\isotope{56}{Ni}$&$\rightarrow$ &$\isotope{56}{Co} + \gamma$ &                                        &8.8 d           &0 -- 18 d\\
                  &             &$\isotope{56}{Co}$         &$\rightarrow \isotope{56}{Fe} + \gamma$ &111.3 d\phantom &18 -- 1100 d  \\
                  &             &                           &$\rightarrow \isotope{56}{Fe} + e^+$    &                &  \\
&&&&&\\
$\isotope{57}{Ni}$&$\rightarrow$ &$\isotope{57}{Co} + \gamma$ &                                        & 2.17 d         &  \\
                  &             &$\isotope{57}{Co}$         &$\rightarrow \isotope{57}{Fe} + \gamma$ & 390 d          &1100 -- 1800 d \\
&&&&&\\
$\isotope{44}{Ti}$&$\rightarrow$ &$\isotope{44}{Sc} + \gamma$ &                                        &87 yrs          &1800 d $\rightarrow$ \\
                  &             &$\isotope{44}{Sc}$         &$\rightarrow \isotope{44}{Ca} + \gamma$ &5.4 h           & \\
                  &             &                           &$\rightarrow  \isotope{44}{Ca} + e^+$   &                &\\
\hline
\end{tabular} 
\end{table}

The gamma-rays produced by the radioactive decay loose their energy by
Compton scattering off bound electrons, producing a population of
non-thermal electrons in the keV range. These, as well as the
positrons, are slowed down by scattering off thermal, free electrons,
ionizations and excitations. This cascade has to be calculated either
by Monte Carlo or by solving the Boltzmann equation, as discussed in
\citet{kf92}. Knowing the heating and ionization rates, one can then
calculate the temperature and ionization state of the gas, and from
this the emission from the different nuclear burning zones. Finally,
by solving the radiative transfer one can calculate a spectrum, which
can be compared to the observations.

\begin{figure} 
\includegraphics*[viewport=0 0 535 351,width=7cm,clip]{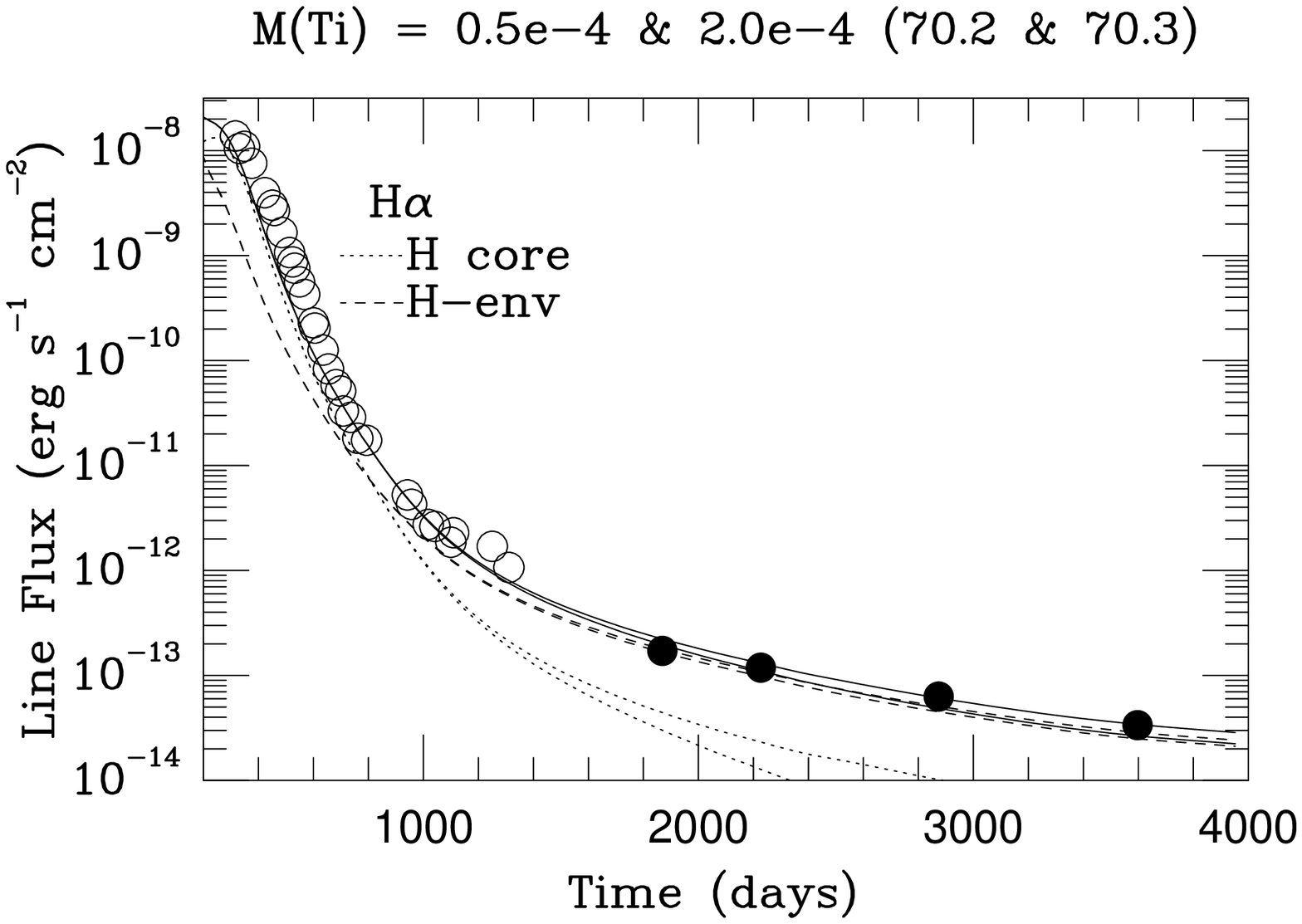}
\includegraphics*[width=7.2cm]{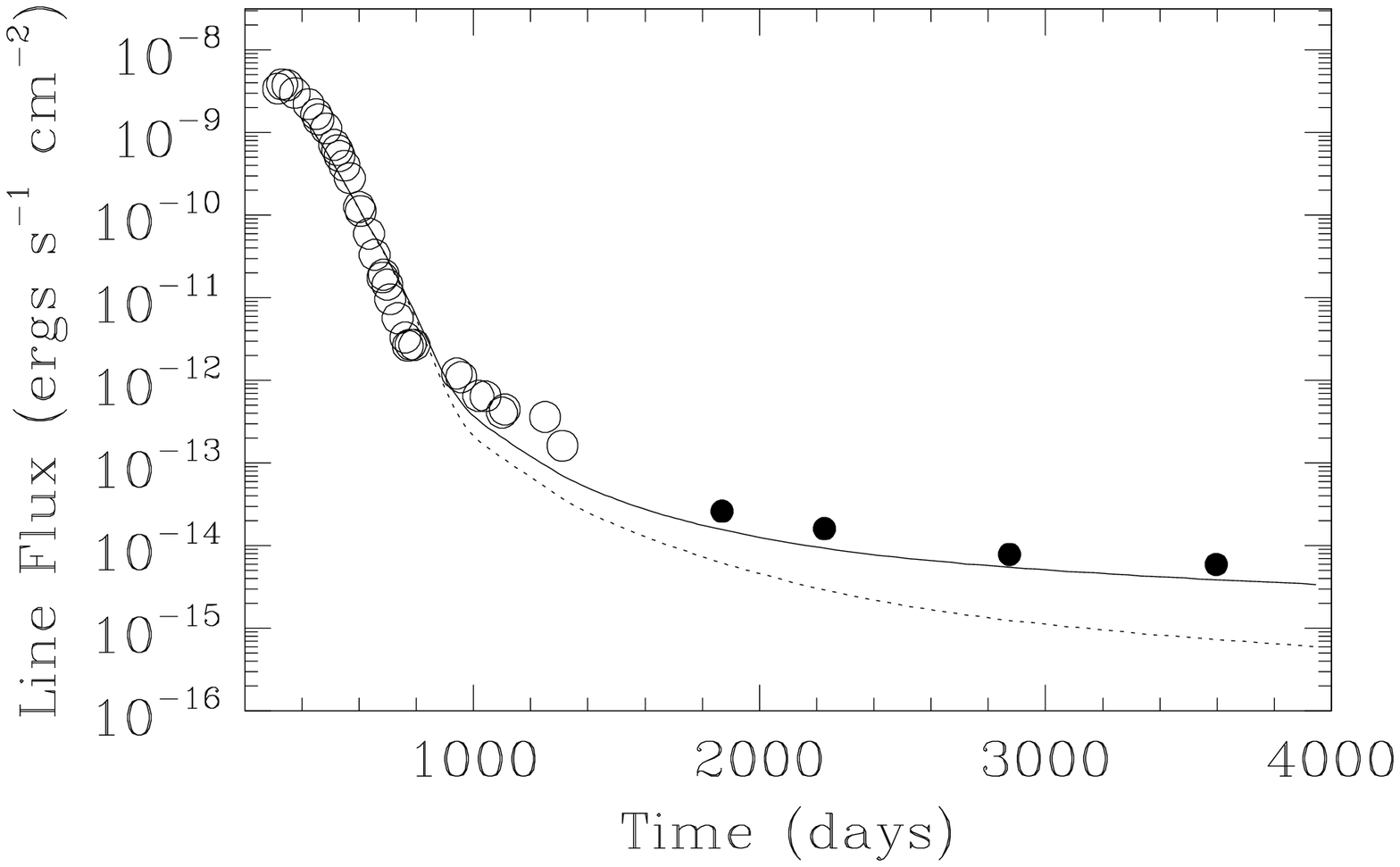}
\caption{Left panel: Evolution of H$\alpha$ from Danziger \etal~ and from SINS/HST
observations together with model calculations. The flat tail is mainly a result of
freeze-out rather than input from ${}^{44}$Ti, as can be seen from the
dominance of the low density envelope (dashed line), compared to the
higher density core (dotted). Right panel: Evolution of the '6300 \AA' line. At early epochs the [O I]
$\wll 6300, 6364$ lines dominate (dotted line), while most of the flux
at epochs $\ga 1400$ days originates from an Fe I line at 6300
\AA. The oxygen mass was in this model $1.9 \Msun$.}
\label{fig1}
\end{figure}

In figure \ref{fig1} we show as an example the light curve of the \Ha~
line compared to data by 
Danziger \etal (1991) and the SINS/HST collaboration (Wang \etal
1996; Chugai \etal 1997). For the explosion
models we have tested both the 11E1 and 14E1 models by  Shigeyama \&
Nomoto(1990) and the 10H
model by Woosley (1988). Thge difference between these models are,
however, marginal. The agreement between the model and
the observations is very good over the whole evolution, which gives
confidence in the whole procedure.  A problem here is that a substantial
fraction of the hydrogen may be at high velocity, where it is
difficult to detect in the faint extended line wings. An accurate
reproduction of the line profile is therefore a must in this case. In
our models we find that the total mass of hydrogen rich gas is $7.7
\pm 2 \Msun$, of which $\sim 2.2 \Msun$ is mixed within 2000
$\kms$. Of the hydrogen rich gas $\sim 3.9 \Msun$ is helium, while an
equal mass is hydrogen.

The mass of the helium dominated zone is consistent with $\sim
1.9\Msun$, of which $\sim
1.8\Msun$ is pure helium, with a further $\sim 3.9 \Msun$ of helium residing in the
hydrogen component. A possible caveat is again that a substantial mass
could be hidden at velocities $\ga ~3000 \kms$.
 
Oxygen is important both as a diagnostic of
the evolutionary calculations, the progenitor mass, and because of its
large abundance.  In figure \ref{fig1}, right panel, we show the corresponding light
curve for this line. A technical point is here that while the feature
at $\sim 6300$ \AA is dominated by the [O I] $\wll 6300, 6364$ doublet
at epochs $\la 1000$ days, it is at late time instead dominated by an
Fe I line at $6300$ \AA. This illustrates the importance of a
self-consistent calculation of the spectrum, taking all
important ions and transitions into account. The total oxygen mass we
find is  $\sim 1.4 \Msun$, but the permissible range is probably
as large as $0.7 - 3.0 \Msun$.  One can from this with good confidence rule out
progenitor models with ZAMS mass less than $15 \Msun$, because of the
low oxygen mass ($\la ~0.5 \Msun$).

In table 2 we summarize the most important masses derived from the spectra and light
curves (bolometric and broad band UBVRI). The latter are discussed in section 3.  Other elements are discussed in \citet{kf98b}.

\begin{table}[t]
\caption{Masses of some important elements and isotopes in SN 1987A.}
\begin{tabular}{lr@{.}lll}
\hline
Isotope&\multicolumn{2}{c}{Mass ($\Msun$)}&Method&Note\\ 
\hline
$\isotope{1}{H}$&3&9&Spectrum&\\
$\isotope{4}{He}$&5&8&Spectrum&\\
$\isotope{16}{O}$&1&9&Spectrum&\\
$\isotope{56}{Ni} ~~(\isotope{56}{Fe})$&0&069&Bol. l.c.&$\pm 0.003 \Msun$\\
$\isotope{57}{Ni}~~(\isotope{57}{Fe})$&0&003&UBVRI l.c.&\\
$\isotope{58}{Ni}$&0&006&Spectrum&\\
$\isotope{44}{Ti}~~(\isotope{44}{Ca})$&0&0001&UBVRI l.c.&$(0.5-2)\EE{-4} \Msun$\\
\hline
\end{tabular} 
\end{table}

\section {The radioactive isotopes and the light curve.}

To estimate the masses of the three most abundant radioactive isotopes
formed in the explosion, \nia, \nib~ and \ti~, one can use several
approaches, each with its special advantages and disadvantages
(Fransson \& Kozma, in preparation). The best known, and for the
\nia~mass most reliable, is the bolometric light curve. As shown by
e.g., \citet{Bouchet_etal_91}, this yields a \nia~ mass of $0.069
\Msun$.

For \nib, or rather, \cob, which dominates the light curve between
1100 -- 1800 days, this is not so straightforward, and a simple
approach similar to that of \coa, yielded a \nib~ mass at least twice
as large as other methods, and much higher than could be produced by
the nucleosynthesis models \citep{Suntzeff et al. (1992),
1991AJ....102.1135B}.  This was shown by \citet{fk93} to be a result
of time dependent effects in the light curve. Because of the
decreasing density and ionization, the recombination time scale
($\propto 1/(\alpha_{rec} n_e)$) as well as the cooling time scale,
become longer than the expansion time scale, $t$. The gas is therefore
not able to recombine and cool at the same rate as the radioactive
input takes place, and only a fraction of the radioactive input is
released as radiation. The difference is released later as
recombination radiation, giving a higher luminosity than the
instantaneous input. The bolometric light curve therefore gives an
overestimate of the radioactive mass if a simple steady state
calculation is employed.

\begin{figure}
\begin{center}
\includegraphics*[width=8cm,angle=-90]{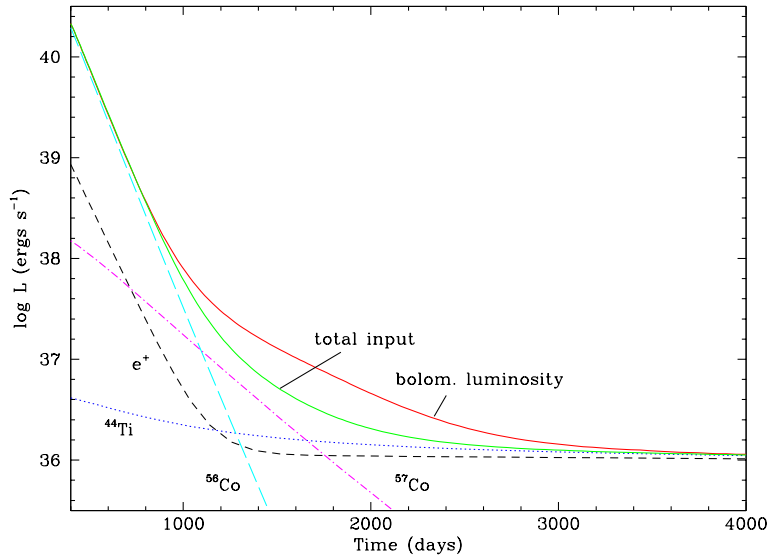}
\end{center}
\caption{Bolometric light curve of SN 1987A with M$({}^{56}$Co$)=0.07 \Msun$,
M$({}^{57}$Co$)=3.3\EE{-3} \Msun$, and M$({}^{44}$Ti$)=1\EE{-4}
\Msun$. The separate contributions from the different isotopes as well
as the positron input are
shown as dashed and dotted lines. Note the dominant positron input
from \ti~ at late epochs.}
\label{fig3}
\end{figure}

In figure \ref{fig3} we show a time dependent calculation of the bolometric light
curve, and also showing the various
individual radioactivity contributions. When we compare the bolometric
luminosity with the total energy input one sees that between $\sim
1200$ days and $\sim 2500$ days the bolometric luminosity is up to a
factor three higher than the input. At very late time most of the energy
input is in the iron rich zones by positrons from \ti, making time
dependent effects less important.  

\begin{figure} 
\begin{center}
\includegraphics*[width=8cm]{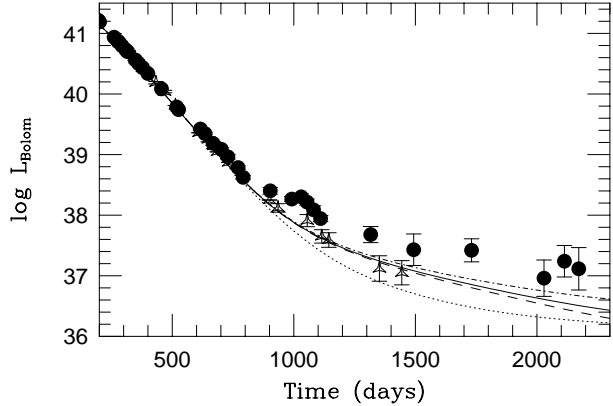}
\end{center}
\caption{Bolometric light curve with  M$({}^{56}$Co$)=0.07 \Msun$,
M$({}^{57}$Co$)=3.3\EE{-3} \Msun$, and M$({}^{44}$Ti$)=1\EE{-4}
\Msun$. Data are from \citet{1993namc.meet..208D} (dots) and from Suntzeff \etal (1992) (triangles). The dotted line
gives the steady state light curve, while the solid shows the same
with time dependent effects included.}
\label{fig4}
\end{figure}

The most serious problem of using the bolometric light curve at late
times is the large bolometric correction necessary because nearly all
of the emission emerges in the unobserved far-IR.  In figure
\ref{fig4} we nevertheless show the time-dependent bolometric light
curve. This represents an extension of the modeling in Fransson \&
Kozma (1993) to later epochs, and with updated physics.  The
freeze-out increases in this case the luminosity by a factor 2 -- 3 at
epochs later than 1200 days.  The observations come from CTIO data by
\citet{Suntzeff et al. (1992)}, and ESO data by
\citet{1993namc.meet..208D}. While the agreement with the CTIO data is
good, the agreement is considerably worse with the ESO data during the
late epochs. The discrepancy between the ESO and CTIO data probably
gives a good estimate of the systematic errors in the observations,
and it is clear that a different approach has to be taken for
estimating the mass of \ti, and also to get a more accurate estimate
of the \nib~ mass.

The second approach involves modeling of the line emission from the
ejecta. An example of this was shown in figure \ref{fig1}, where we
show \Ha~ light curves for three different values of the \ti~
mass. . Although the model reproduces the observations very well
during the whole evolution, as a measure of the ${}^{44}$Ti mass the
light curve is clearly not very sensitive.  Nearly all of the
excitation is due to gamma-rays absorbed by the envelope, and the line
is therefore strongly affected by the freeze-out. Most of the emission
at $\la ~3000$ days is in fact due to the delayed recombination from
the epoch at $\la ~2000$ days when ${}^{57}$Co dominated the energy
input. A better alternative is to use lines from the metal rich core,
in particular the Fe I and Fe II lines.  An important result found by
Chugai \etal (1997) is that in order to reproduce the Fe II flux,
positrons have to be trapped efficiently in the iron rich gas. Pure
Coulomb collisions are not sufficient for this, but a weak magnetic
field is enough to make the gyro radius of the positrons small
enough. A severe problem is, however, that modeling of individual
lines is very sensitive to atomic data, which in spite of dramatic
improvements from the IRON-project, is still a problem. In a similar
attempt \citet{Lundqvist_et_al01} have tried to use the FIR
[Fe II] $26.0 \mu$ fine structure line in connection with observations by
ISO. Unfortunately, only an upper limit to the flux was obtained,
yielding an upper limit to the \ti~ mass of $1.1\EE{-4} \Msun$. There
is, however, a number of loop holes in this result, such as extra dust
cooling, mixing with other heavy elements contributing to the
cooling like silicon and sulphur, as well as uncertainties in the
atomic data.

A compromise between these two approaches is offered by broad band
photometry, where photometry in the optical and near-IR is compared to
synthetic photometry from spectral calculations. This method avoids
the need to extrapolate into the far-IR for the bolometric light
curve, as well as many of the uncertainties involved in the
calculation of individual lines. Rather than modeling individual
lines, the broad band fluxes are mainly sensitive to the division
between UV, optical and IR in the spectral distributions of the
elements. This is determined by the atomic configuration,
i.e. the term diagram, and less by the exact path a recombining photon
takes when cascading to the ground state. Except for the fraction of
the emission in the optical and IR which is absorbed by the dust, the
photometry is insensitive to this component. The fraction absorbed by
dust can be estimated from mid-IR observations, as well as from line
profile shifts (Lucy \etal 1991).

\begin{figure} 
\begin{center}
\includegraphics*[width=10cm]{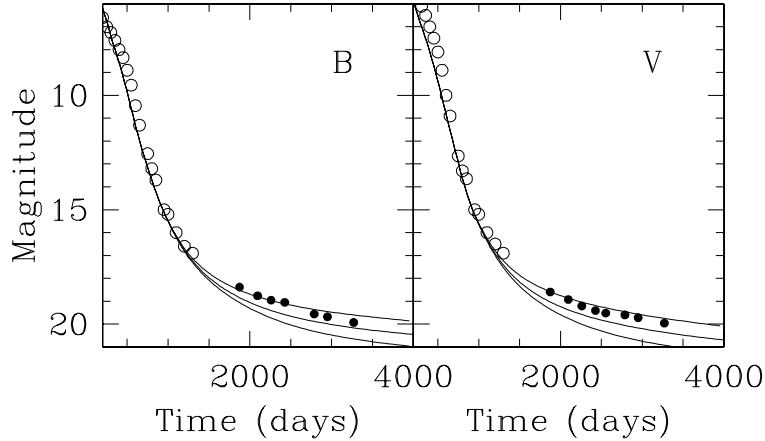}
\end{center}
\caption{B and V light curves from Suntzeff \etal (1991, 2001),
together with models with $(0.5, 1.0, 2.0)\EE{-4} \Msun$ of
${}^{44}$Ti. All models have $3.3\EE{-3} \Msun$ of ${}^{57}$Ni}
\label{fig5}
\end{figure}

In figure \ref{fig5} we show the B and V band light curves together
with observations. The reason for concentrating on these bands is that
they are strongly dominated by lines from Fe I and Fe II, and
therefore most sensitive to the \ti~ mass, in contrast to the R band
which is dominated by \Ha, and therefore sensitive to the freeze
out. Observations are taken from Suntzeff \etal (1991) and HST
observations from Suntzeff \etal (2001).  The models have $6.9\EE{-2}
\Msun$ of ${}^{56}$Co, $3.3\EE{-3} \Msun$ of ${}^{57}$Co and three
different values of the \ti~ mass, $(0.5, 1.0, 2.0)\EE{-4} \Msun$. As
can be seen, both the B and V bands are well reproduced by the
observations throughout the whole evolution, and the currently best
estimate of the \ti~mass is in the range $(0.5-2.0)\EE{-4}
\Msun$. This method also yields the best estimate of the \nib~ mass,
$\sim 3.3\EE{-3} \Msun$. A more complete discussion of these results
will be given in Fransson \& Kozma (2001).

\section{Conclusions}
SN 1987A is by far the best source of nucleosynthesis of any
SN. Reliable masses for both the most abundant elements, H, He, and O,
as well as the the most abundant radioactive isotopes, \nia, \nib, and
\ti~ have been obtained. These provide some of the most important
constraints on the progenitor and on the explosion itself. 

CF is grateful to Roland Diehl for organizing a very successful
workshop, and to Don Clayton and Nikos Prantzos for useful discussions
and comments.



\begin{thebibliography}{}

\bibitem[Bouchet \& Danziger(1993)]{1993A&A...273..451B} Bouchet, P.~\& 
Danziger, I.~J.\ 1993, \aap, 273, 451 

\bibitem[Bouchet, Danziger, \& Lucy(1991)]{1991AJ....102.1135B} Bouchet, 
P., Danziger, I.~J., \& Lucy, L.~B.\ 1991, \aj, 102, 1135 


\bibitem[Bouchet et al.(1991)]{Bouchet_etal_91} Bouchet, P., Phillips,
M.~M., Suntzeff, N.~B., Gouiffes, C., Hanuschik, R.~W., \& Wooden,
D.~H.\ 1991, \aap, 245, 490

\bibitem[Chugai, Chevalier, Kirshner, \& 
Challis(1997)]{chugai97} Chugai, N.~N., Chevalier, R.~A., 
Kirshner, R.~P., \& Challis, P.~M.\ 1997, \apj, 483, 925 

\bibitem[Danziger \& Bouchet(1993)]{1993namc.meet..208D} Danziger, I.~J.~\& 
Bouchet, P.\ 1993, New Aspects of Magellanic Cloud Research, 208 

\bibitem[Danziger et al. (1991)] {Danziger et al. (1991)} Danziger, I.~J.,  Bouchet, P., Gouiffes, C., Lucy,
L.~B. 1991, in Proc. ESO/EIPC
Supernova Workshop,  SN 1987A and other
Supernovae, eds. I.J. Danziger \& K. Kj\"ar, (Garching: ESO),  217

\bibitem[de Kool, Li, \& McCray(1998)]{deKool98} de 
Kool, M., Li, H., \& McCray, R.\ 1998, \apj, 503, 857 

\bibitem[Fransson (1994)] {Fransson (1994)} Fransson, C., 1994, in  Supernovae (Les Houches, Session
LIV 1990), eds. J. Audouze, S. Bludman, R. Mochkovitch, \& J. Zinn-Justin 
(New York: Elsevier),  677

\bibitem[Fransson \& Kozma(1993)]{fk93} Fransson, C.~\& 
Kozma, C.\ 1993, \apjl, 408, L25 

\bibitem[Kifonidis, Plewa, Janka, \& M{\" 
u}ller(2000)]{2000ApJ...531L.123K} Kifonidis, K., Plewa, T., Janka, H.-T., 
\& M{\" u}ller, E.\ 2000, \apjl, 531, L123 

\bibitem[Kozma \& Fransson(1992)]{kf92} Kozma, C.~\& 
Fransson, C.\ 1992, \apj, 390, 602 

\bibitem[Kozma \& Fransson(1998a)]{kf98a} Kozma, C.~\& 
Fransson, C.\ 1998, \apj, 496, 946 

\bibitem[Kozma \& Fransson(1998b)]{kf98b} Kozma, C.~\& 
Fransson, C.\ 1998, \apj, 497, 431 

\bibitem[Kumagai et al.(1993)]{Kumagai et al. (1993)} Kumagai, S., Nomoto, 
K., Shigeyama, T., Hashimoto, M., \& Itoh, M.\ 1993, \aap, 273, 153 

\bibitem[Lucy, Danziger, Gouiffes, and Bouchet (1991)] {Lucy et
al. (1991)} Lucy, L.B., Danziger, I.J., Gouiffes, C., and Bouchet,
P. 1991, in Supernovae, Proc. of the Tenth Santa Cruz Summer Workshop
in Astronomy and Astrophysics, ed. S.E. Woosley (Springer Verlag), 82.

\bibitem[Lundqvist et al. (2001)] {Lundqvist_et_al01} Lundqvist, P.,
Kozma, C., Sollerman, J. \& Fransson, C. 2001, \aap, 374, 629

\bibitem[Phillips et al. (1990)] {Phillips et al. (1990)} Phillips, M. M., Hamuy, M., Heathcote, S. R., Suntzeff, N. B.,
\& Kirhakos, S. 1990, \aj, 99, 1133

\bibitem[Shigeyama \& Nomoto(1990)]{Shigeyama and Nomoto (1990)} Shigeyama, 
T.~\& Nomoto, K.\ 1990, \apj, 360, 242 

\bibitem[Suntzeff et al.(1991)]{Suntzeff et al. (1991)} Suntzeff, N.~B., 
Phillips, M.~M., Depoy, D.~L., Elias, J.~H., \& Walker, A.~R.\ 1991, \aj, 
102, 1118 

\bibitem[Suntzeff et al.(1992)]{Suntzeff et al. (1992)} Suntzeff, N.~B., 
Phillips, M.~M., Elias, J.~H., Walker, A.~R., \& Depoy, D.~L.\ 1992, \apjl, 
384, L33 

\bibitem[Suntzeff et al.(2001)]{Suntzeff et al. (2001)} Suntzeff, N.~B. 
et. al. 2001, in preparation


\bibitem[Timmes, Woosley, Hartmann, \& Hoffman(1996)]{Timmes et al. (1996)} 
Timmes, F.~X., Woosley, S.~E., Hartmann, D.~H., \& Hoffman, R.~D.\ 1996, 
\apj, 464, 332 

\bibitem[Wang et al.(1996)]{Wang et al. (1996)} Wang, L.~et al.\ 1996, \apj, 
466, 998 

\bibitem[Woosley (1988)]{Woosley (1998)} Woosley, S.~E. 1988, \apj, 330, 218

\end{thebibliography}
\end{document}